\documentclass[%
 reprint,
 amsmath,amssymb,
 aps,
]{revtex4-2}

\usepackage{graphicx}
\usepackage{dcolumn}
\usepackage{bm}

\begin{document}

\preprint{APS/123-QED}

\title{Charge-Density-Waves Tuned by Crystal Symmetry}



\author{A. Gallo-Frantz}
\author{A.A. Sinchenko}
\author{D. Ghoneim}
\author{L. Ortega}
\author{V.L.R. Jacques}
\email{vincent.jacques@universite-paris-saclay.fr}
\author{D. Le Bolloc'h}
\email{david.le-bolloch@universite-paris-saclay.fr}
 \affiliation{Laboratoire de Physique des Solides, Université Paris-Saclay, CNRS, 91405 Orsay Cedex, France}

\author{P. Godard}
\author{P.-O. Renault}
\affiliation{Institut Pprime, CNRS-Université de Poitiers-ENSMA, 86962 Futuroscope-Chasseneuil Cedex, France}

\author{P. Grigoriev}
\affiliation{L. D. Landau Institute for Theoretical Physics, Chernogolovka, Moscow Region 142432, Russia}

\author{A. Hadj-Azzem}
\author{P. Monceau}
\affiliation{Univ. Grenoble Alpes, CNRS, Grenoble INP, Institut Néel, 38000 Grenoble, France}

\author{D. Thiaudière}
\affiliation{Synchrotron SOLEIL, L'Orme des Merisiers, 91190 Saint-Aubin, France}

\author{E. Bellec}
\affiliation{European Synchrotron Radiation Facility, 71 Avenue des Martyrs, 38043 Grenoble Cedex 9, France}

\date{\today}

\begin{abstract}
The electronic orders appearing in condensed matter systems are originating from the precise arrangement of atoms constituting the crystal as well as their nature. This teneous relationship can lead to highly different phases in condensed matter, and drive electronic phase transitions. Here, we show that a very slight deformation of the crystal structure of TbTe$_3$ can have a dramatic influence on the electronic order that is stabilized. In particular, we show that the Charge Density Wave (CDW) developping along the $\vec{c}$ axis in the pristine state, switches to an orientation along $\vec{a}$ when the naturally orthorhombic system is turned into a tetragonal system. This is achieved by performing true biaxial mechanical deformation of a TbTe$_3$ sample from 250K to 375K, and by measuring both structural and electronic parameters with x-ray diffraction and transport measurements. We show that this switching transition is driven by the tetragonality parameter $a/c$, and that the transition occurs for $a=c$, with a coexistence region for $0.9985< a/c < 1.002$. The CDW transition temperature $T_c$ is found to have a linear dependence with $a/c$, with no saturation in the deformed states investigated here, while the gap saturates out of the coexistence region. The linear dependence of $T_c$ is accounted for within a tight-binding model. Our results question the relationship between the gap and $T_c$ in RTe$_3$ systems. More generally, our method of applying true biaxial deformation at cryogenic temperatures can be applied to many systems displaying electronic phase transitions, and opens a new route towards the study of coexisting or competing electronic orders in condensed matter. 
\end{abstract}

\maketitle


\section{\label{sec:intro}Introduction}
 The electronic properties of materials are intimately linked to their atomic structure. The latter drives phase transitions such as metal-to-insulator transitions through electron-phonon coupling or by modification of the electron hoping energy from one site to another. This coupling to the lattice induces such phases as the Mott insulating state, charge density waves (CDW) and superconductivity (SC). CDW, that have been studied for decades, mainly in low-dimensional systems~\cite{Gruner1994a, Monceau2012}, were found to appear near the SC phase of cuprates~\cite{tranquada1995}, which raised an increasing amount of new studies about the competition between CDW and SC. As they are very sensitive to electron-phonon coupling, application of strain is key to tune electronic properties, and in a more general way to better understand those complex electronic orders. The methods to play on structural parameters are diverse (physical and chemical pressure, epitaxial strain in thin films) and cover many fields of condensed matter physics. In recent years, application of direct mechanical deformation using piezoelectric actuators~\cite{shayegan2003} has been applied successfully in quantum materials to drive electronic transitions. This technique is very powerful to induce structural anisotropies and is particularly well suited to study quasi-2D systems in which most electronic correlations are taking place within specific atomic planes of the crystal, and which are weakly coupled to the other planes in the perpendicular direction. Elastoresistivity measurements performed under uniaxial stress allowed for instance to study nematic susceptibility in iron pnictides~\cite{fisher2012, fisher2013, eckberg2020, frachet2022} or heavy fermion materials~\cite{fisher2015}. However, although compatible with cryogenic temperatures, devices applying uniaxial stress have an intrinsic limitation in terms of flexibility of the mechanical strains that can be achieved in the samples, in particular because when the sample is deformed along a chosen direction, the transverse ones also deform by a certain amount which is determined by the intrinsic Poisson ratio of the material under consideration. The structural parameters thus cannot be modified at will, as only one direction of strain is controlled, the other ones following their own mechanical behaviour accordingly. It is for instance impossible to explore a case for which two directions of the crystal are elongated simultaneously, to get an homothetic transformation of the parameters in two directions of the crystal. In addition, the deformation value of the sample along all crystallographic directions is generally not directly measured. 
 
 Here we present results obtained by true biaxial mechanical deformation of quasi-2D materials at cryogenic temperatures, allowing to probe both electronic and structural parameters, using both resistivity and x-ray diffraction (XRD) measurements. To do so, a specific device has been developed~\footnote{A paper describing the device is in preparation.}, and its principle is further described in section~\ref{sec:ExpDesc}. 

\begin{figure*}[!ht]
\includegraphics[width = 17cm]{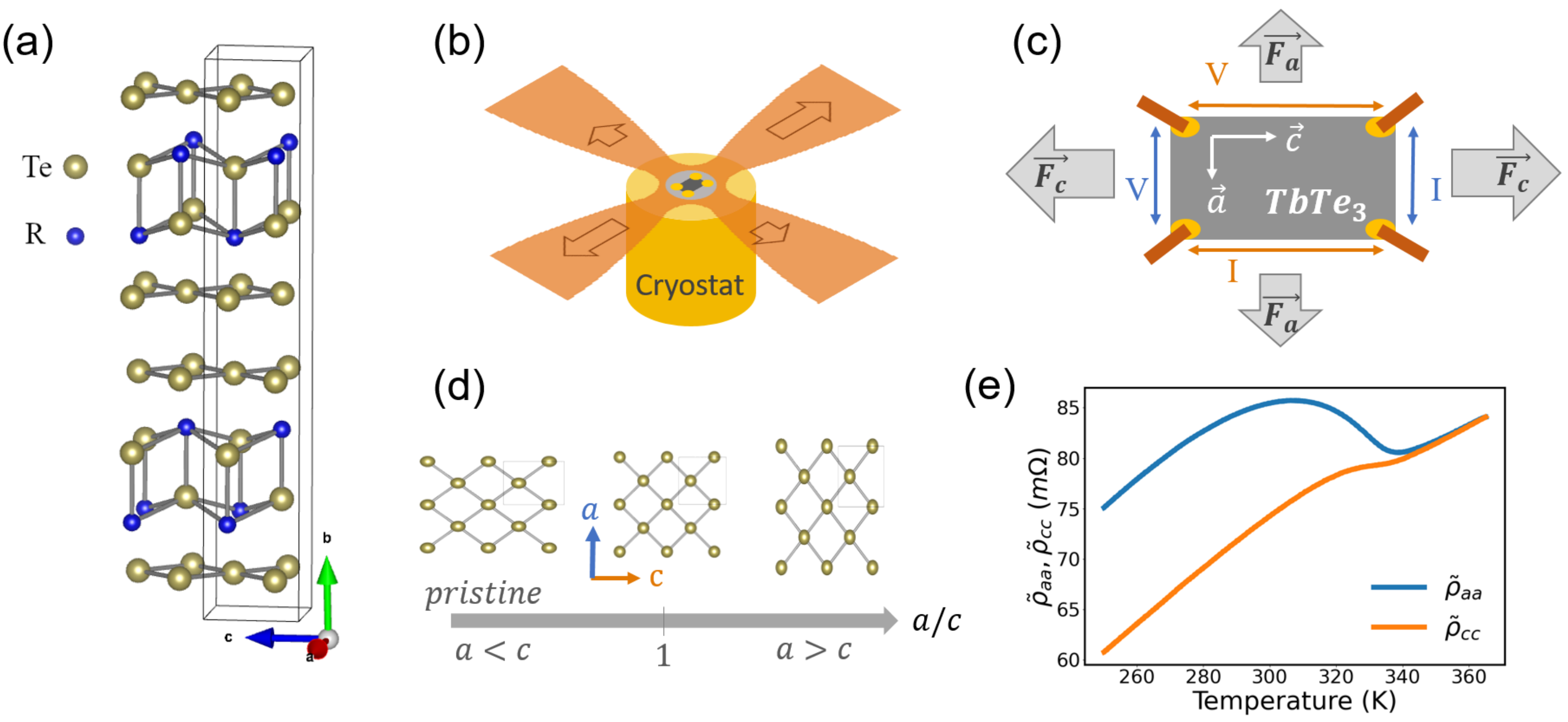}
\caption{\label{fig:setup}(a) Crystal structure of RTe$_3$ systems, with the quasi-tetragonal unit cell shown in black lines. A view of the nearly-square Te sheet in the $\left(\vec{a},\vec{c}\right)$ plane is shown on the side. (b) Schematic drawing of the biaxial tensile stress device: a thin crystalline sample is glued on a polyimide cross-shaped substrate on which tensile stresses are applied along the branches. The main in-plane directions of the crystal $\vec{a}$ and $\vec{c}$ are aligned with the branches of the cross. The bottom side of the cross is in contact with the cold finger of a nitrogen flow cryostat. (c) Four electrical contacts are deposited at the four corners of the sample, allowing to apply currents and measure voltages along $\vec{a}$ or $\vec{c}$ directions of the crystal. The forces applied along the opposite branches of the cross have same magnitude, both along $\vec{a}$ and $\vec{c}$ ($\vec{F_a}$ and $\vec{F_c}$ respectively). (d) Schematic view of crystal deformation as a function of a/c structural parameter. (e) Resistivities along $a$ and $c$ divided by sample thickness as a function of temperature. }
\end{figure*}
 
 We focus on TbTe$_3$, one the RTe$_3$ compounds, where $R$ is a rare earth element of the Lanthanide family ($R$ = La, Ce, Pr, Nd, Sm, Gd, Tb, Dy, Ho, Er, Tm)~\cite{yumigeta2021}. The crystalline structure of this material is quasi-tetragonal, and made of a succession of RTe slab intercalated by two quasi-square Te planes in the $\left(\vec{a},\vec{c}\right)$ planes (see Fig.~\ref{fig:setup}(a)). The two in-plane directions are non-equivalent due to the presence of a glide plane along $\vec{c}$ (space group $Cmcm$) which leads to a slight orthorhombicity ($1-\frac{a}{c}\sim 1.3\cdot 10^{-3}$ in TbTe$_3$ at 300K \cite{Ru2008}). These systems present a highly rich phase diagram, with a CDW appearing along the c direction in all RTe$_3$ systems at high temperature ($T<T_c$) with wavevector $\vec{Q_{c}}=\left(0,0,\sim\frac{2}{7}c^*\right)$, and a second one at lower temperature ($T<T_{c_2}$) with wavevector $\vec{Q_{a}}=\left(\sim\frac{2}{7}a^*,0,0\right)$ for the heavier R elements (R = Tb, Dy, Ho, Er and Tm)~\cite{Ru2008}. Both CDW are incommensurate with the underlying lattice period, as demonstrated by the continuous temperature evolution of $Q_{c}$ and $Q_{a}$ and by non-linear transport properties~\cite{sinchenko2012,sinchenko2016}. In addition, a magnetic phase appears at low temperature for the heaviest R as well as SC under pressure at temperatures $\sim$1K for some compounds in the series~\cite{Ru2008a, zocco2015}.
 
 The temperature dependence of the CDW transitions as a function of R is generally attributed to a chemical pressure effect, the lattice parameters evolving continuously with the size of the rare-earth element~\cite{diMasi1995}. Hydrostatic physical pressure also shifts the CDW transition temperatures in qualitatively the same way as chemical pressure \textit{i.e.} towards lower temperatures for $T_c$ and higher temperatures for $T_{c_2}$~\cite{zocco2015, sacchetti2009}. Recently, elastoresistivity and elastocaloric measurements performed under uniaxial stress in  ErTe$_3$ and TmTe$_3$ suggested a possible switching of the CDW wavevector orientation from $\vec{c}$ to $\vec{a}$~\cite{straquadine2022}. A significant change of $T_{c_2}$ was also reported when the sample is deformed along $\vec{a}$, while only a very slight change of $T_c$ was observed. However, in the latter study, the change of $a$ and $c$ lattice parameters could not be measured or changed simultaneaously and resistivity measurements are only performed along the applied stress direction, which prevents to get the full information on both CDW in all experimental conditions. 

 \section{Experimental details}
 \label{sec:ExpDesc} 
Here, we use biaxial in-plane tensile stress to apply controlled deformation along the two in-plane crystallographic axis $\vec{a}$ and $\vec{c}$ of TbTe$_3$. Single crystals were grown by the self-flux method, as described in ~\cite{sinchenko2012} and glued at the center of a 125µm-thick polyimide cross-shaped substrate, with $\vec{a}$ and $\vec{c}$ in-plane directions aligned with the branches of the polyimide cross (see Fig.~\ref{fig:setup}(b)). The four branches of this substrate are attached to four independent motors that can pull on each branch separately. The forces applied along the four branches are recorded using calibrated force gauges, and are given in kg. The center of the cross is covered with a thin gold layer on its bottom surface and lies on the cold finger of a nitrogen-flow cryostat, allowing to reach temperatures in the range 80-350K. Apiezon grease is used to get a good thermal transfer between the cold finger and the polyimide cross. In practice, the same motor displacement is used for opposite branches to keep the sample at the same position in the device. This whole setup is enclosed in a vacuum chamber for cryogenic operation, and allows access for incoming and outgoing x-rays to perform x-ray diffraction in reflection geometry on the sample through a 300µm-thick Polyether-ether-ketone dome. 

As the deformation of the substrate is transferred to the sample through its glued surface, the sample is mechanically exfoliated down to a thickness $d \sim$2.5$\mu$m thickness to get homogeneous deformation in the volume. The sample shape is rectangular in the $(\vec{a}, \vec{c})$ plane, with a 1:1.17 aspect ratio (with $\sim$1 mm lateral dimensions). Four contact are deposited at the four corners of the crystal, and the Montgomery method is used to get both $R_{aa}$ and $R_{cc}$ resistances along the $a$ and $c$ directions respectively by switching the direction of measurement~\cite{Montgomery1971, Logan1971, ong1978, dosSantos2011}. The ratio of transverse dimensions of the sample $L_a/L_c = 1.17$ was obtained by fixing the anisotropy to 1 in the normal state and used to compute the resistivities normalized by sample thickness $\tilde\rho_{aa}=\rho_{aa}/d$ and $\tilde\rho_{cc}=\rho_{cc}/d$. More information on resistivity measurements and switching procedure are provided in Supp. Mat. In this way, stresses can be applied and resistivities can be measured along both $\vec{a}$ and $\vec{c}$ directions at will (see Fig.~\ref{fig:setup}(c)). This also allowed us to directly determine the crystal orientation as  $\rho_{aa}$ and $\rho_{cc}$ have a clearly different behaviour through $T_c$~\cite{sinchenko2014}. The resistivities $\tilde\rho_{aa}$ and $\tilde\rho_{cc}$ obtained in the pristine state between 250K and 365K are presented in Fig.~\ref{fig:setup}(e).


$\tilde\rho_{aa}$ and $\tilde\rho_{cc}$ display the typical behaviour observed in RTe$_3$ compounds, following a linear behaviour in the normal state, above the CDW transition, and presenting a resistivity jump below $T_{c}~\sim337\pm 1K$ when entering the CDW phase. $\tilde\rho_{aa}$ is particularly sensitive to the transition towards the CDW along $\vec{c}$ due to the Fermi surface shape. Indeed, the gap opening annihilates electronic states with velocities along $\vec{a}$ which consequently induces an increase of $\rho_{aa}$ by $\sim 2\mu\Omega \cdot cm$. On the contrary, $\rho_{cc}$ only presents a short plateau at $T_{c}$. Due to this feature, the appearance of CDW along $\vec{a}$ or $\vec{c}$ is better seen in the resistivity curves measured perpendicular to the CDW wavevector direction. 

In the following, we present results obtained when applying tensile stresses along the two in-plane directions $\vec{a}$ and $\vec{c}$ independently, as well as along both directions simultaneously, in equibiaxial mode, \textit{i.e.} using the same stress for the two orthogonal directions, which is equivalent to in-plane biaxial negative pressure. Resistivity and XRD measurements were both performed during in-situ tensile deformation of the sample. 

XRD measurements were performed to follow the structural changes during deformation, as well as the satellite peaks associated to the CDW. The experiment was performed with a 8keV x-ray beam generated by a Cu rotating anode source (Rigaku RU-300B). The sample mounted in the biaxial tensile stress device was positioned at the center of rotation of a Huber eulerian 4-circle diffractometer to perform wide-angle x-ray diffraction. Three non-colinear Bragg reflections (0 16 0, 0 16 1 and 1 15 0) were measured with a 2D detector (Timepix from ASI) located 82cm after the sample to retrieve the three lattice parameters at all forces $F_a$ and $F_c$ applied along $\vec{a}$ and $\vec{c}$ respectively. The absence of twin domains was demonstrated by checking that the forbidden 0 15 1 reflection was indeed not measurable.

\section{Results}
\label{sec:Results}

\begin{figure}[!ht]
\includegraphics[width = \columnwidth]{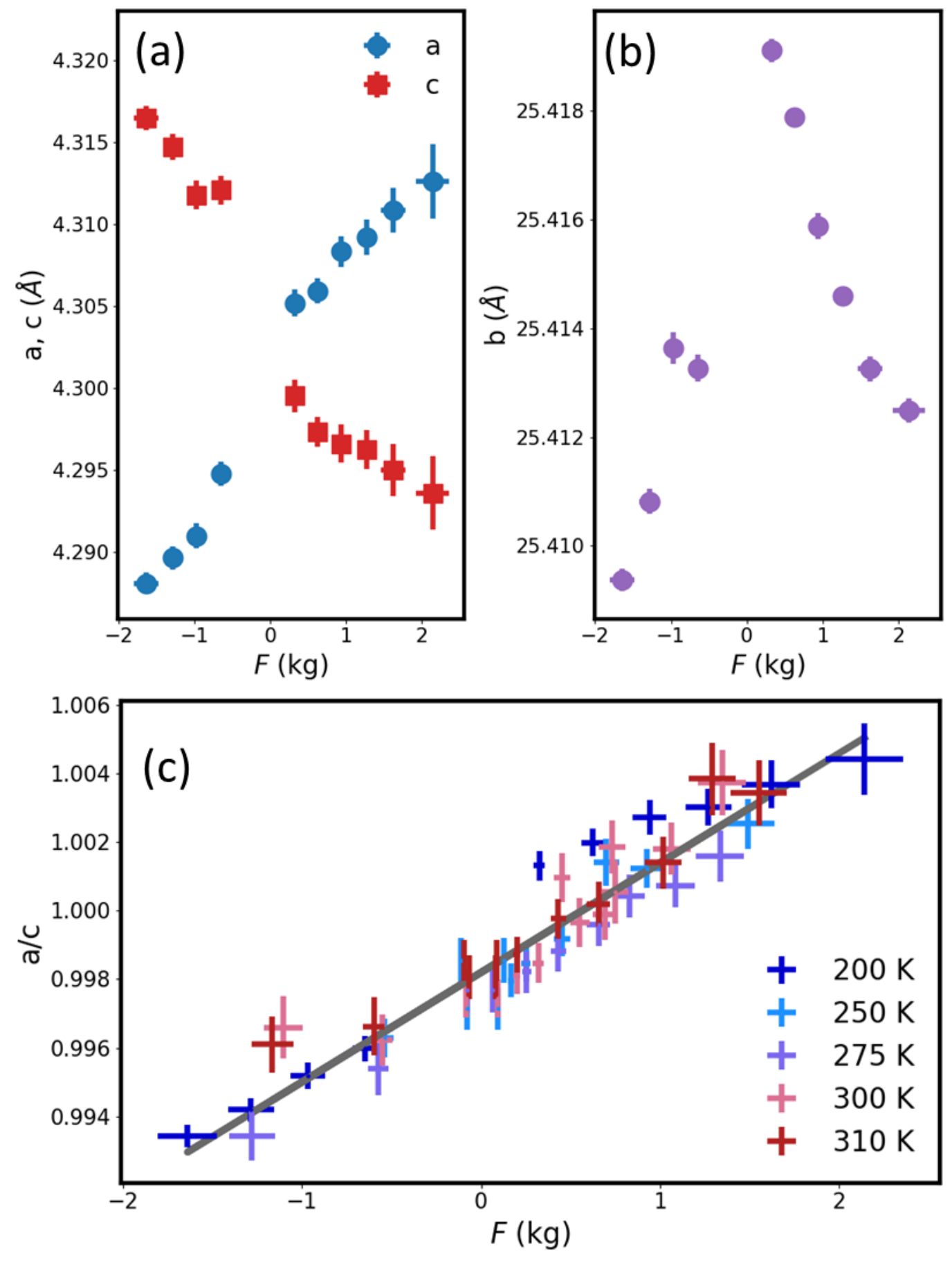}
\caption{\label{fig:a_c_a-over-c} Evolution of (a) in-plane and (b) out-of-plane lattice constants of TbTe$_3$, obtained from the 3 colinear Bragg peaks 0 16 0, 1 15 0 and 0 16 1, as a function of applied forces along $F_{a}$ and $-F_{c}$. This representation allows to see that the relative variations of a and c are correlated with a Poisson ratio close to 1, which indicates that the tensile stress along one of the 2 in-plane directions is equivalent to uniaxial pressure along the other one. (c) Evolution of the a/c ratio as a function of the same forces, at several temperatures between 200K and 310K. The grey line is obtained by a linear adjustment of all data points. Without applied force, a/c $\sim 0.998 \pm 0.001$, and $a/c=1$ is obtained for $F_a\sim0.7\pm0.1$ kg.}
\end{figure}

The 3 Bragg reflections recorded on the 2D detector were projected along the $2\theta$ direction of reciprocal space to compute the three lattice parameters $a$, $b$ and $c$ for each set of applied forces. Their evolution is plotted as a function of uniaxial forces $F_a$ and $-F_c$ in Fig.~\ref{fig:a_c_a-over-c}(a) and (b). 
$a$ and $c$ follow a quasi-linear behaviour when applying a uniaxial force $F_a$, with a relative variation $\Delta a/a \sim 0.3\%$ and  $\Delta c/c \sim -0.3\%$ for $F_a = 2$ kg. The deformations observed on $a$ and $c$ are similar for forces $F_c$. The in-plane Poisson ratio is thus found close to 1. The $b$ lattice parameter decreases both when applying forces $F_a$ and $F_c$, but with relative variations which are ten time smaller ($\Delta b/b \sim 0.03\%$) which correponds to a Poisson ratio of 0.1 in this direction. This strong difference can be explained by the weak van der Waals coupling between layers along $\vec{b}$. The evolution of the $a/c$ ratio has been plotted as a function of uniaxial forces in Fig~\ref{fig:a_c_a-over-c}(c) for several temperatures. Interestingly, this ratio evolves linearly as a function of applied uniaxial force, with $a/c = A\times F + B$, where $F=F_a$ when $F>0$ and $F=-F_c$ when $F<0$, and $A = 0.0032$ kg$^{-1}$ and $B = 0.9982$. The value found for $B$ is consistent with the expected $a/c$ ratio in the pristine state. The direct correlation of uniaxial forces to the tetragonality parameter $a/c$ will be used to express all relevant quantities in terms of $a/c$ in the following.

\begin{figure}[!ht]
\includegraphics[width = \columnwidth]{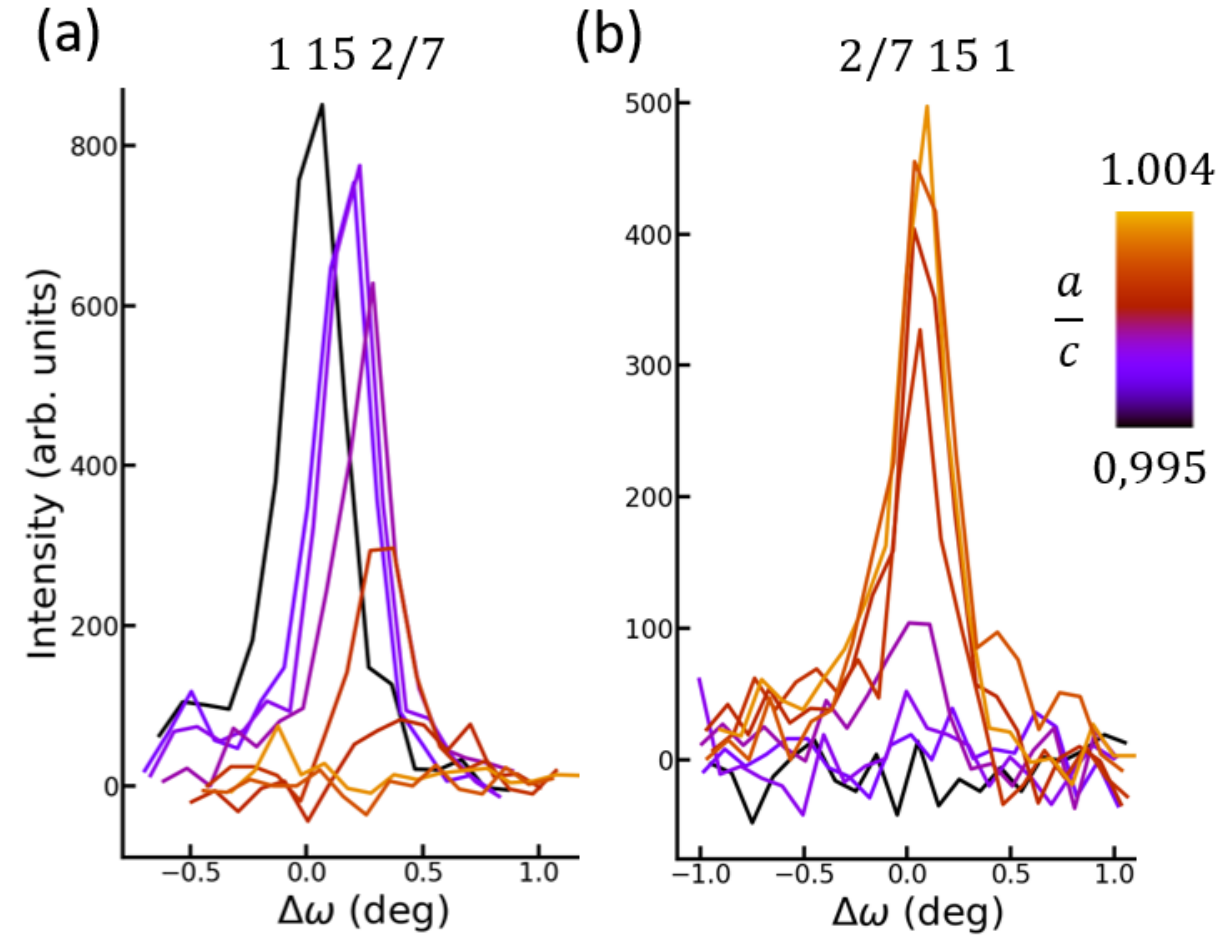}
\caption{\label{fig:XRD_Satellites} Rocking scans on the (a) 1 15 2/7 and (b) 2/7 15 1 peaks associated to the CDW along $\vec{c}$ and $\vec{a}$ respectively as a function of a/c ratio, at T=250K. The a/c ratio was computed from the uniaxial forces $F_a$ and $F_a$ using the linear expression described in the text. The rocking angle $\omega$ is taken relative to the peak position obtained for the lowest a/c value.}
\end{figure}

The CDW can also be tracked as a function of applied force by XRD. Indeed, this new periodicity in the system leads to the appearance of additional reflections around lattice Bragg reflections. We measured the 1 15 $\sim2/7$ satellite reflection associated to the CDW along $\vec{c}$ to track the changes induced by mechanical deformation on this CDW component, and the $\sim$2/7 15 1 associated to the CDW along $\vec{a}$ as it was shown to appear transiently after laser excitation~\cite{kogar2020} and suggested that it could appear under mechanical stress~\cite{straquadine2022}. The evolution of these two satellites was followed as a function of applied uniaxial forces $F_a$ and $F_c$, at T=250K. The corresponding rocking curves are shown in Fig.~\ref{fig:XRD_Satellites}(a) and (b) as a function of $a/c$ that was computed from $F_a$ and $F_c$ using the linear expression mentioned previously.

The purple curves are the ones obtained in the pristine state. The 1 15 2/7 is a single peak, which accounts for a single-domain CDW, with a rocking width $\sim 0.5$ deg. The 2/7 15 0 is completely absent, which again confirms the absence of twin domains in the pristine state. When a force $F_c$ is applied, $a/c$ decreases and the 1 15 2/7 peak intensity increases with no clear change of width, while the 2/7 15 1 is still absent. This is true down to the lowest $a/c$ value reached here ($\sim 0.995$). On the contrary, when a force $F_a$ is applied along $\vec{a}$, the $a/c$ ratio increases, and the 1 15 2/7 peak intensity decreases until it disappears completely at the highest $a/c=1.004$. At the same time, the 2/7 15 1 peak increases until it reaches $\sim5/8$ of the maximum intensity of the 1 15 2/7. Its width is similar to the 1 15 2/7, \textit{i.e.} $\sim0.5$ deg. This measurement indicates that when $a/c$ increases, the CDW along $\vec{c}$ progressively disappears while at the same time the CDW appears along $\vec{a}$ with a coexistence region where both are present. 

\begin{figure*}[!ht]
\includegraphics[width = 17cm]{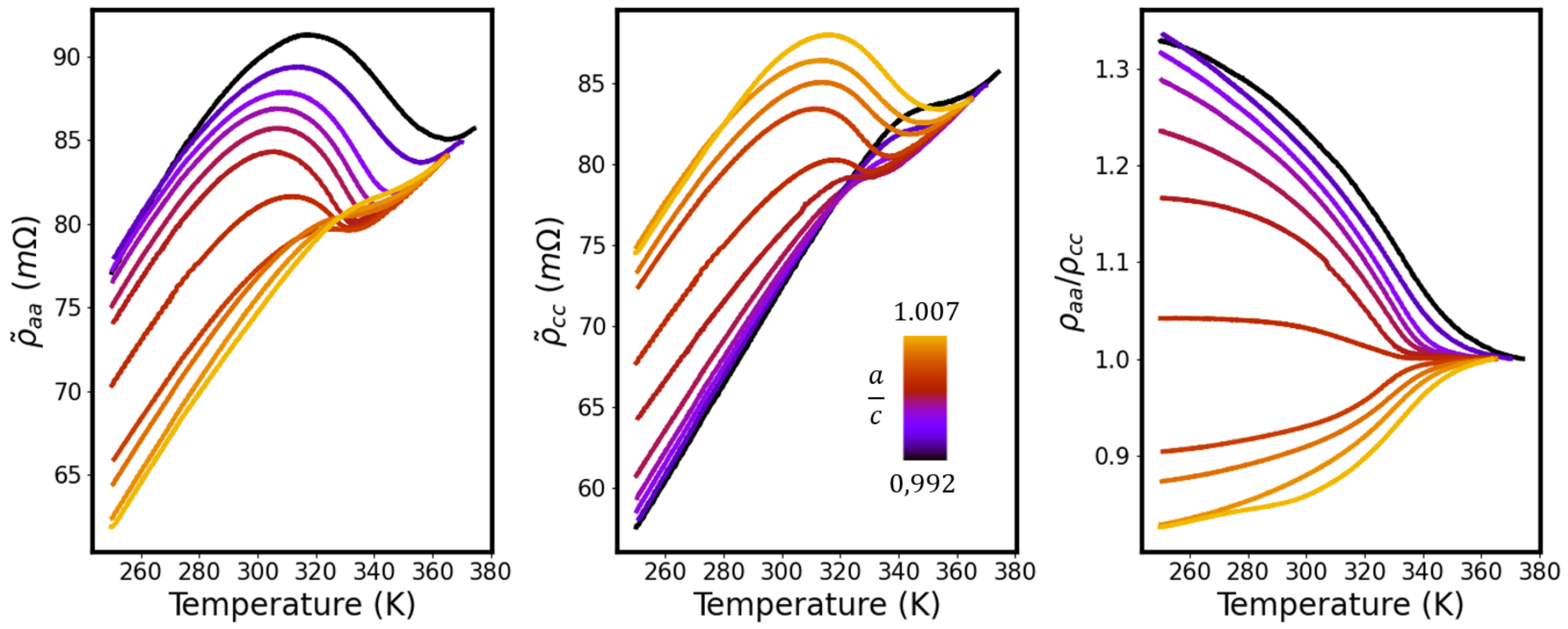}
\caption{\label{fig:rhoaa_rhocc_anisotropy}(a)-(b) $\tilde\rho_{aa}$ and $\tilde\rho_{cc}$ resistivity curves obtained along $\vec{a}$ and $\vec{c}$ respectively between 250K and 375K, as a function of $a/c$, varied between 0.992 and 1.007. (c) Anisotropy  $\rho_{aa}/\rho_{cc}$ in the same temperature range, obtained from the curves shown in (a) and (b).} 
\end{figure*}

Resistivity measurements were performed in the same sample, after measuring the XRD data. $\rho_{aa}$ and $\rho_{cc}$ were recorded at several uniaxial forces between 250K and 375K, well below and above $T_{c}$. Again, the uniaxial forces $F_a$ and $F_c$ were converted into $a/c$ ratio to get a continuous picture of the resistivity evolution around the tetragonality point. $\rho_{aa}$, $\rho_{cc}$ and the anisotropy $\rho_{aa}/\rho_{cc}$ are plotted in Fig~\ref{fig:rhoaa_rhocc_anisotropy}. 

In the pristine state (purple curve in Fig.~\ref{fig:rhoaa_rhocc_anisotropy}), as plotted in Fig~\ref{fig:setup}(e), $\rho_{aa}$ and $\rho_{cc}$ display the usual behaviour, and the anisotropy, which tends to 1 for $T>T_c$, increases below $T_c$ and reaches 1.3 at 250K, which is consistent with the usual anisotropy value measured at this temperature in TbTe$_3$~\cite{sinchenko2014}. When $a/c$ decreases, $\rho_{aa}$ shifts towards higher resistivity values, by $\sim1\mu\Omega\cdot$cm for $a/c=0.992$, but also towards higher temperatures, and keeps the same shape as in the pristine state. Simultaneously, $\rho_{cc}$ behaves the same, shifting to higher resistivity and higher temperatures by the same amount as $\rho_{aa}$, but keeping the same shape as in the pristine state. The anisotropy mainly shifts in temperature by the same amount, and $T_c$ is found to shift by $\sim$20K in this state. The behaviour when $a/c$ increases is more complex. In this case, the amplitude of $\rho_{aa}$ decreases while the amplitude of $\rho_{cc}$ increases until $\rho_{aa}$ has a similar shape than $\rho_{cc}$ in the pristine state and inversely $\rho_{cc}$ has a similar shape than $\rho_{aa}$ in the pristine state (yellow curves in Fig.~\ref{fig:rhoaa_rhocc_anisotropy})(a) and (b)). We thus get a continuous inversion of $\rho_{aa}$ into $\rho_{cc}$ and inversely. In addition, the evolution of $T_c$, that can be tracked qualitatively by following the position of the local minimum of the resistivity curves between 320K and 375K, first decreases and then increases again. The anisotropy clearly reveals the inversion of $\rho_{aa}$ and $\rho_{cc}$, as when $a/c$ increases the anisotropy in the CDW state decreases, and goes below 1 to reach $\sim$0.825 at 250K for $a/c=1.007$.

\begin{figure}[!ht]
\includegraphics[width = 0.8\columnwidth]{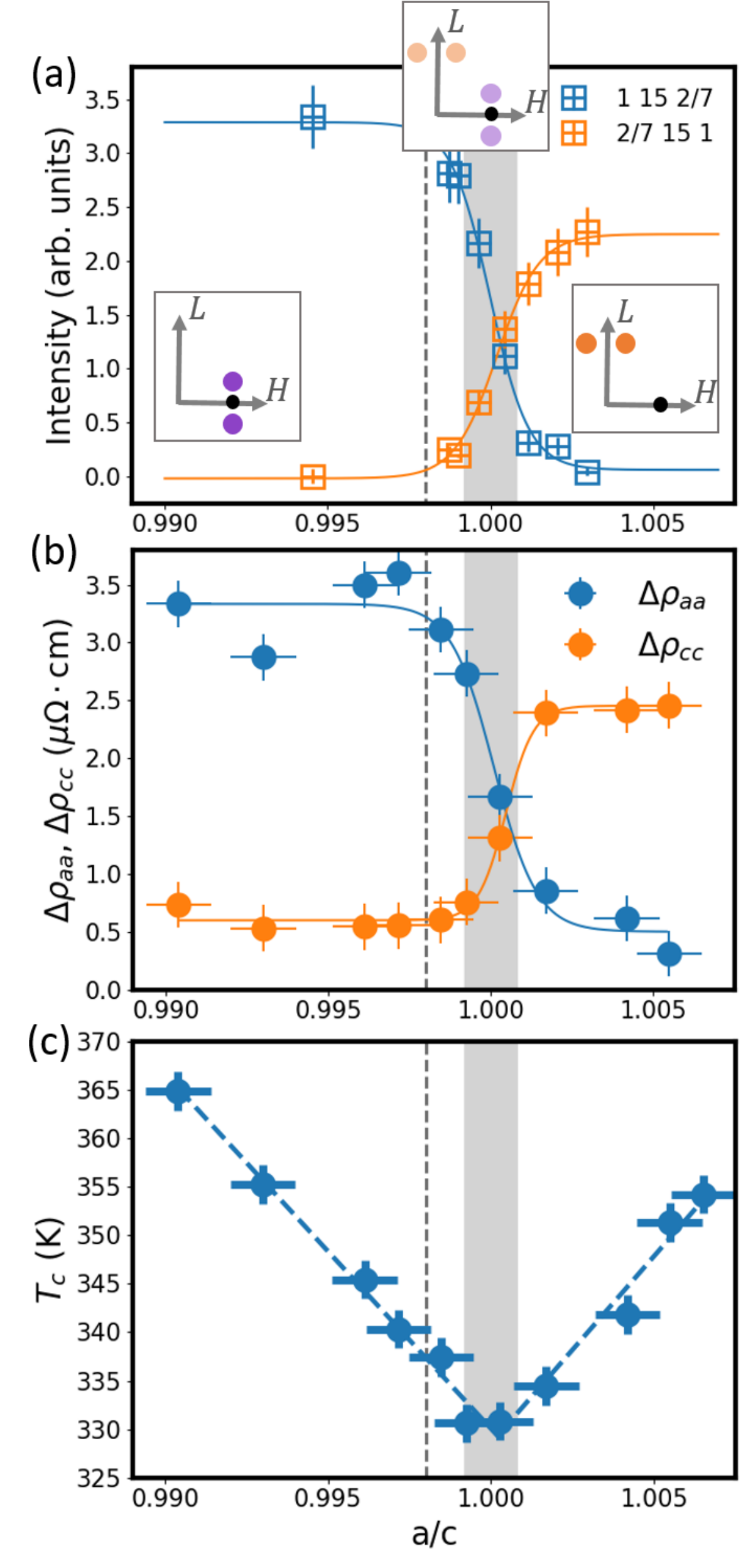}
\caption{\label{fig:Dintensity_Drho_Tc}(a) Evolution of the 1 15 2/7 (blue squares) and 2/7 15 1 (orange squares) satellite intensities measured as a function of $a/c$. The blue (resp. orange) solid line is an inverse-sigmoid (resp. sigmoid) fit to the experimental points obtained on the 1 15 2/7 (resp. 2/7 15 1) satellite reflection. The three sketches in the inset represent a (H 15 L) plane in reciprocal space. The black dot is the 1 15 0 Bragg reflection, and the purple (resp. orange) dots are the 1 15 $\pm$2/7 (resp. $\pm$2/7 15 1) reflections, with a color strength depending on the measured intensity in each $a/c$ region. (b) Resistivity jumps $\Delta\rho_{aa}$ (blue dots) and $\Delta\rho_{cc}$ (orange dots) extracted from the experimental curves shown in Fig.~\ref{fig:rhoaa_rhocc_anisotropy} using the method described in the text, plotted as a function of $a/c$. The blue (orange) solid line is an inverse-sigmoid (resp. sigmoid) fit of $\Delta\rho_{aa}$ (resp. $\Delta\rho_{cc}$). (c) $T_c$ computed from the resistivity curve taking the average of the local peaks found in the second derivative of $\rho_{aa}$, $\rho_{cc}$ and $\rho_{aa}/\rho_{cc}$ as a function of $a/c$. For all three panels, the dashed gray line is the $a/c$ value in the pristine state, and the gray filled region marks the position $a/c=1$ with a width equal to the error bar ($\pm0.8\cdot10^{-3}$).}
\end{figure}

The evolution of both XRD and resistivity measurements are plotted as a function of the tetragonality parameter $a/c$ in Fig~\ref{fig:Dintensity_Drho_Tc}. 
More specifically, in Fig~\ref{fig:Dintensity_Drho_Tc}(a), we computed the integrated intensities of the 1 15 2/7 and 2/7 15 1 satellite reflections shown in Fig~\ref{fig:XRD_Satellites}. AS previously described, the intensity of the 1 15 2/7 decreases while 2/7 15 1 increases when $a/c$ increases from 0.994 to 1.004, and follow a sigmoid shape for the 2/7 15 1 and an inverse-sigmoid shape for 1 15 2/7, having the same width, and both centered at the same position $a/c=1$ (within the error bars). The saturation value of the 2/7 15 1 is $\sim$2/3 that of the saturation value of the 1 15 2/7, and both go down to zero when they reach their minimum value. When $a/c<0.999$, we only measure a satellite along $\vec{c}$ and when $a/c>1.002$, the satellite along $\vec{c}$ has completely disappeared, and the satellite along $\vec{a}$ is maximum. We thus observe a continuous transformation of the CDW along $\vec{c}$ into a CDW along $\vec{a}$ when increasing the $a/c$ ratio, and an intermediate coexistence phase for $0.999<a/c<1.002$, and a crossing point at $a/c=1$. 

The same kind a plot has been performed for the resistivity jump $\Delta\rho$ obtained from $\rho_{aa}$ and $\rho_{cc}$ by taking the maximum value of $\rho-\rho_n$, where $\rho_n$ is the linear resistivity in the normal state ($T>T_c$). The values $\Delta\rho_{aa}$ and $\Delta\rho_{cc}$ obtained with this method are shown in Fig~\ref{fig:Dintensity_Drho_Tc}(b). The shape of both curves is identical to the ones obtained for the satellite intensities in Fig.~\ref{fig:Dintensity_Drho_Tc}(a): $\Delta\rho_{aa}$ has an inverse-sigmoid shape and $\Delta\rho_{cc}$ a sigmoid shape, with the same width as the ones obtained for the satellites in XRD. They also cross at the same point $a/c=1$, with an intermediate region in the range $0.999<a/c<1.002$. Also, similarly to the satellite intensity curves, the saturation value obtained for $\Delta\rho_{cc}$ is $\sim2/3$ the saturation value of $\Delta\rho_{aa}$. 

Finally, a value of $T_c$ has been extracted from the resistivity curves. $T_c$ was determined by taking the second derivative of the resistivity curves and spotting the local peak corresponding to the inflection point of the resistivities. By doing so, the $T_c$ obtained from $\rho_{aa}$, $\rho_{cc}$ and anisotropies were consistent. The evolution of $T_c$ is plotted as a function of $a/c$ in Fig.~\ref{fig:Dintensity_Drho_Tc}(c). $T_c$ decreases linearly for $a/c<1$, and then increases linearly again for $a/c>1$, with a minimum value of 330K found at $a/c=1$. $T_c$ was increased by 30K compared to the pristine state, but could potentially go on increasing to much larger values if the $a/c$ parameter could be explored in a wider range of values. The linear dependence is consistent with the behaviour reported in ~\cite{straquadine2022}, but with a much greater amplitude. 

\begin{figure}[!ht]
\includegraphics[width = \columnwidth]{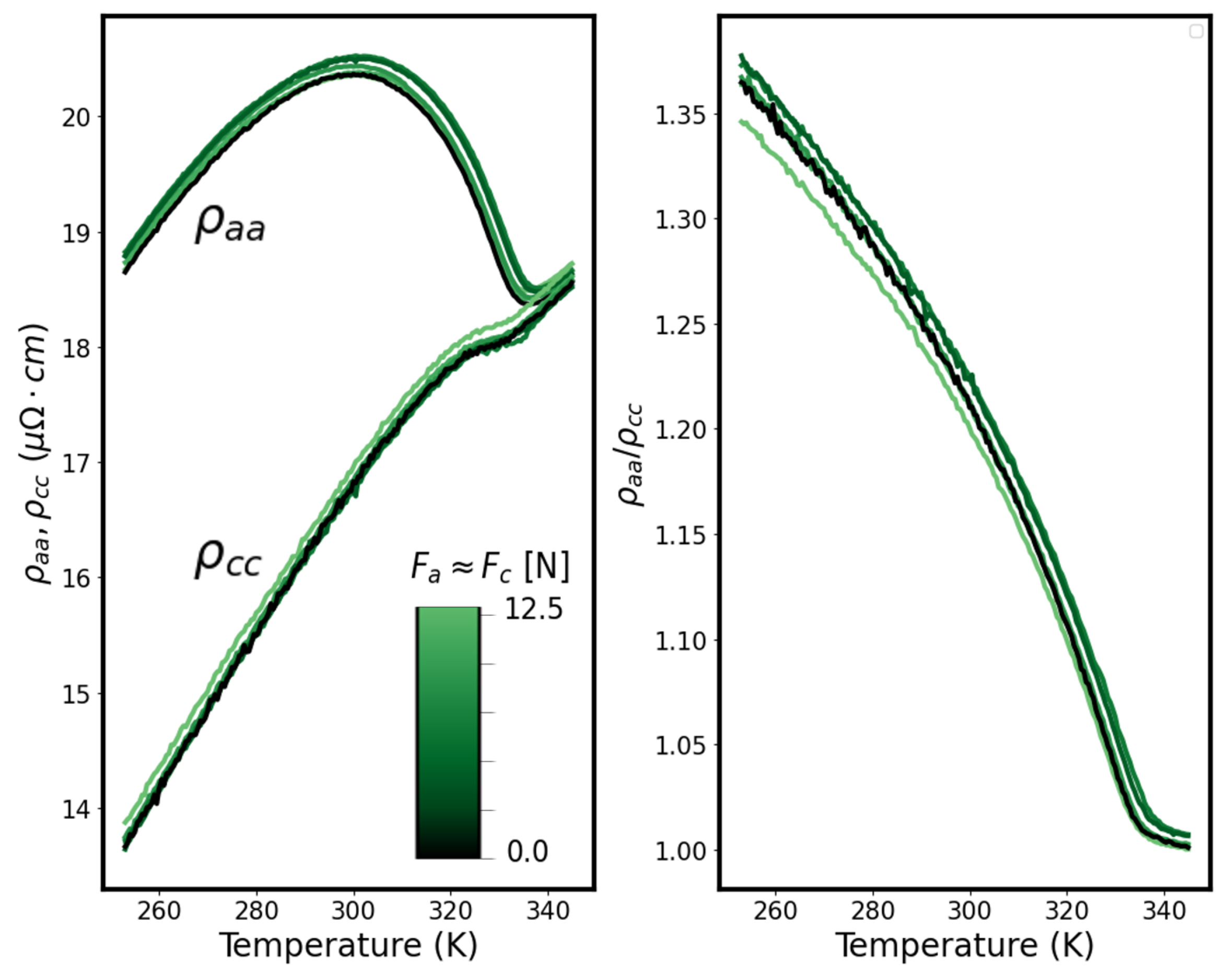}
\caption{\label{fig:rhoaa_rhocc_equibiaxial} (a) $\rho_{aa}$ and $\rho_{cc}$ measured in equibiaxial mode, with $F_a=F_c$, measured between 250K and 350K, up to 1.25kg on both axes. (b) $\rho_{aa}/\rho_{cc}$ computed from the resistivities shown in (a).}
\end{figure}

A final measurement supporting the fact that the $a/c$ ratio is the key parameter in the CDW switching from $\vec{c}$ to $\vec{a}$ orientation was performed by applying an equibiaxial deformation to the sample. In this case, the same forces were applied along both $a$ and $c$ directions of the sample, and the resisitivity were followed as a function of the force. The resistivities and anisotropies obtained in these conditions are shown in Fig.~\ref{fig:rhoaa_rhocc_equibiaxial}. 

Although the forces where increased along both directions up to 1.25kg, \textit{i.e.} the same force range as the ones used in the uniaxial case, $\rho_{aa}$ and $\rho_{cc}$ keep the same shape and position. Only slight changes of resistivity values are observed, which can be attributed to the precision on the applied forces. Changing the absolute values of $a$ and $c$ keeping the $a/c$ ratio constant in an equibiaxial tensile stress experiment thus does not affect the pristine CDW state. 

\section{Discussion}
\label{sec:Disc}
All the results presented above show that the key parameter for the CDW orientational switching from $\vec{c}$ to $\vec{a}$ is driven by the tetragonality parameter $a/c$. This result is of prime importance as it shows the direct relationship between the crystal symmetry and the appearance of CDWs in TbTe$_3$, and presumably in all RTe$_3$ systems as a hint of this transition was also reported in TmTe$_3$ and ErTe$_3$ in~\cite{straquadine2022}. It is also notable that the glide plane does not play any role in the CDW stabilisation as it is still present during crystal deformation - the forbidden reflection 0 15 1 does not appear during the experiment. When the crystal symmetry is tetragonal, \textit{i.e.} when $a=c$, both CDW, with orientation along $\vec{c}$ and $\vec{a}$ coexist, as revealed by XRD and transport measurements. It should be emphasized here that those results are extremely reproducible, both when changing the $a/c$ ratio on a sample from one phase to another, and also from one sample to another. 

In XRD experiments, the intensity of the satellite reflections associated to the CDW is generally related to the amplitude of the periodic lattice distorsion (PLD) associated to the CDW, which is related to the order parameter and is in general proportional to the CDW gap squared. In addition, its position relative to the Bragg peak position in reciprocal space directly gives the evolution of the CDW wavevector. Here, the intensity of the satellite associated to the CDW along $\vec{c}$ decreases when $a/c$ increases, while the intensity of the satellite associated to the CDW along $\vec{a}$ increases. This suggests that the amplitude of the PLD along $\vec{c}$ vanishes while the the one along $\vec{a}$ appears, and that the gap must close along $\vec{c}$ and appear along $\vec{a}$. 

This is consistent with the resistivity measurements. Indeed, the amplitude of the jumps $\Delta\rho_{aa}$ and $\Delta\rho_{cc}$ are related to the gap along $\vec{c}$ and $\vec{a}$ respectively. As shown in the results, this is consistent with a gap closure along $\vec{c}$ when $a/c$ increases while it opens along $\vec{a}$, with a mixed state in the region $a=c$. The inversion of the anisotropy, that is also associated to the gap, is also a proof of this gap switching. This should appear in ARPES experiments upon tensile deformation of the sample. 

The clear coexistence region when $0.999 < a/c < 1.002$ demonstrates that a mixed state can be stabilized, with gaps in both directions. It should be investigated whether they appear in separate domains in the sample, or in different Te planes, or if they are superimposed in the same Te planes.  

The present results can be compared to the behaviour observed in hydrostatic pressure experiments or with chemical pressure obtained by changing the rare-earth element involved. Applying hydrostatic pressure was shown to be equivalent to chemical pressure~\cite{sacchetti2009}: when the lattice parameters are compressed, the CDW wavevector, the gap and $T_c$ evolve similarly in both cases. However, here, when $a$ and $c$ are increased similarly, no change in resistivity is observed, meaning that the gap and the $T_c$ do not change in the negative pressure region. To observe variations of gap and $T_c$, the tretragonality parameter $a/c$ has to be changed. In the experiments reported in~\cite{sacchetti2009}, the parameter $a/c$ chnages and tends to 1 when pressure is applied (above 3GPa in CeTe$_3$). In these conditions, $T_c$ decreases, which is in qualitative agreement with our observations: $T_c$ is minimum when $a/c=1$. A quantitative comparison is more difficult, but the variation of $T_c$ reported in~\cite{sacchetti2009} are much larger than the ones reported here. Concerning the evolution of the gap, they also report a vanishing of the CDW satellite intensities, suggesting a vanishing of the gap at high pressures, while $a/c=1$. Here, no change is observed in equibiaxial mode, but the experiments should be repeated in equibiaxial conditions while $a/c=1$. The role of the relative variation of the $b$ parameter should also be investigated further to compare our experiment to hydrostatic pressure measurements, although the inter-plane coupling is much weaker due to the van der Waals nature of the bonds in this direction. 

The behaviour observed here is very similar to the one observed after fs laser excitation reported in~\cite{kogar2020}, with the CDW appearing along $a$ while the one along $c$ disappears. Contrary to the equilibrium CDWs, the CDWs appearing along $a$ and $c$ in this reference and in our study cannot coexist with one another. This makes the new CDW appearing along $a$ unique and different from the equilibrium CDW found at low temperatures along $a$ that coexists with the one along $c$. Following the results presented in~\cite{kogar2020}, it could be interesting to investigate further the lattice parameter evolution of LaTe$_3$ after laser pulse excitation and check whether the system becomes transiently tetragonal or displays a transient evolution of $a/c$ similar to the one measures here. In our case, both CDW are stable as long as the $a/c$ parameter is kept constant. 

Finally, the most striking feature is the evolution of $T_c$ compared to the evolution of the gap, as extracted from XRD and resistivity measurements. Both satellite intensities and $\Delta\rho$ saturate when $a/c<0.9985$ and $a/c>1.002$, meaning that the gap value saturates in this region. However, $T_c$ keeps diverging linearly in these regions, which questions the link between the gap and $T_c$. The linear dependence of $T_c$ with respect to $a/c$ can be understood within the tight-binding model though. In this description, the electron dispersion of the two bands formed by the $p_x$ and $p_y$ Te orbitals $\varepsilon_{\pm}$ keeps its form:

\begin{equation}
    \varepsilon_{\pm} = -2t_{\parallel}\cos{\left(k_x^* \pm k_y^*\right)} - 2t_{\perp}\cos{\left(k_x^* \mp k_y^*\right)}
\label{eq:epsilons}
\end{equation}

where $k_x^* = k_xa/2$ and $k_y^* = k_yc/2$, and $t_{\parallel}$ and $t_{\perp}$ are the transfer integrals parallel and perpendicular to the orbital direction. Both lattice parameters and transfer integrals depend on the applied stress. At $a=c$, Eq.~\ref{eq:epsilons} coincides with the accepted tight-binding dispersion given by Eq.(1) in ~\cite{Brouet2008}. The CDW transition temperature is given by the condition :

\begin{equation}
    U\left(\vec{Q_0}\right)\chi\left(T_c,\vec{Q_0}\right) = 1
    \label{eq:U(Q)chi(Q)}
\end{equation}

where the static electron-electron interaction $U\left(\vec{Q}\right)$ inculdes both Coulomb and phonon-mediated interaction, and the Lindhard susceptibility reads :

\begin{equation}
    \chi\left(T_c,\vec{Q}\right) = \sum_{\alpha, \alpha'}\sum_{k_x, k_y}16\frac{n_F\left(E_{\vec{k},\alpha}\right) - n_F\left(E_{\vec{k}+\vec{Q},\alpha'}\right)}{E_{\vec{k}+\vec{Q},\alpha'} - E_{\vec{k},\alpha}}
    \label{eq:Lindhard}
\end{equation}

where $n_F\left(\varepsilon\right) = 1/(1 + exp\left[\left(\varepsilon - E_F\right)/T\right])$ is the Fermi-Dirac distribution function, $\alpha,\alpha'=\pm$ numerate the subbands and $E_{\vec{k},\alpha}$ differs from Eq~\ref{eq:epsilons} only near the intersection points of two bands in momentum space, as given by Eq.(2) in~\cite{Grigoriev2019}. As $ \varepsilon_{\pm}$ and $E_{\vec{k},\pm}$ depend only on $k_x^*$ and $k_y^*$, and as $\chi\left(T_c,\vec{Q}\right)$ has a $x \leftrightarrow y$ symmetry, an equibiaxial stress does not affect the Lindhard susceptibility. The main difference in the transition temperatures of the CDW along $\vec{a}$ and $\vec{c}$ thus comes from the difference of $U\left(\vec{Q}\right)$ when $a \neq c$. The electron-electron coupling mainly comes from the Coulomb interaction, screened by conducting electrons: $U\left(r\right) = e^2 exp\left(-\zeta r\right)/r$, where the inverse Debye screening radius $\zeta = \sqrt{4\pi e^2\rho_F}$ and $\rho_F$ is the density of states at the Fermi level. In TbTe$_3$, $\zeta \approx a^{-1} \approx \left(4.3\right)^{-1}$ \AA$^{-1}$. The Fourier transform of the screened Coulomb potential is 

\begin{equation}
    U\left(\vec{Q}\right) \approx \frac{4\pi e^2}{\vec{Q}^2+\zeta^2}
    \label{eq:U(Q)}
\end{equation}

The CDW wavevector $\vec{Q_0}$ is given by the maximum of the Lindhard susceptibility~\ref{eq:Lindhard}, and gives a fixed product $Q_{0x}a = Q_{0y}c = 10\pi/7$, when $a=c$. An increase of the lattice constant $a$ decreases the CDW wavevector $Q_{0x}$ and increases the CDW coupling $U\left(\vec{Q_0}\right)$ according to Eq.~\ref{eq:U(Q)}. An increase of $a$ by $\delta a/a = 0.1\% $ results in a decrease of $Q_{0x}^2$ by 0.2$\%$. Since $\zeta<<Q_{0x}$, according to Eq.~\ref{eq:U(Q)}, this gives an increase of $U\left(\vec{Q_0}\right)$ by 0.2$\%$. According to Eq.~\ref{eq:U(Q)chi(Q)}, an increase $\delta U$ of $U\left(\vec{Q_0}\right)$ raises the CDW transition temperature $T_c$ by $\delta T_c$ given by Eq.~\ref{eq:U(Q)chi(Q)}: 

\begin{equation}
    \frac{\delta U}{U} = \frac{- \delta \chi}{\chi} = \frac{d\chi\left(T,\vec{Q_0}\right)}{dT} \frac{-\delta T_c}{\chi\left(T,\vec{Q_0}\right)}
\end{equation}

because the electron susceptibility $\chi\left(T,\vec{Q_0}\right)$, approximately given by Eq.~\ref{eq:Lindhard} decreases with the increase of $T$. The temperature dependence of the Lindhard susceptibility was calculated iné\cite{Grigoriev2019} for the second CDW in ErTe$_3$ and iné\cite{vorobyev2019} for the Q-dependence of the first CDW for various parameters of electron dispersion. To find $d\chi/dT$ for the first CDW in relevant temperature range and for the transfer integrals $t_{\parallel}\approx2eV$, $t_{\perp}\approx0.37eV$ and the Fermi energy $E_F\approx1.48eV$ in TbTe$_3$~\cite{Brouet2008}, we performed new calculations of $\chi\left(T,\vec{Q_0}\right)$ using Eq.~\ref{eq:Lindhard} (see Supp. Mat.), from which the slope of the temperature-dependent susceptibility can be extracted :

\begin{equation}
    \eta = \frac{d\left[ln \chi\left(T, \vec{Q_0}\right)\right]}{dT} = \chi^{-1}\frac{d\chi}{dT} \approx -1.5\cdot10^{-4} K^{-1}
\end{equation}

We thus can compute $\delta T_c$ for an increase $\delta a/a \approx 0.2\%$, as was experimentally done here: 

\begin{equation}
    \delta T_c = \frac{\delta U\left(Q_0\right)}{\eta U\left(Q_0\right)} \approx -\frac{2 \delta a}{a \eta} \approx \frac{4\cdot10^{-3}}{1.5\cdot10^{-4}} \approx 26K
    \label{eq:Tccomputed}
\end{equation}

In our experiment, a 0.2$\%$ increase of the lattice constant $a$ results in an increase $\delta T_c \approx 20K$, which is consistent with the value computed in Eq.~\ref{eq:Tccomputed}. The same applies to deformations along $c$, and explains the linear dependence of $T_c$ as a function of deformation. 

The system actually chooses the CDW along the axis with the highest $T_c$. For $a>c$, $T_{cx}>T_{cy}$ and the CDW is aligned along $a$, while for $a<c$, $T_{cx}<T_{cy}$ and the CDW is aligned along $c$. At $a=c$, the transition temperatures are equal $T_{cx}=T_{cy}$, and the system is degenerate. A spatial phase segregation between CDW aligned along $a$ and $c$ may appear, accompanied by stronger fluctuations.

\section{Conclusion}
\label{sec:ccls}
In this article, we used \textit{in-situ} x-ray diffraction and transport to follow the structural and electronic properties of the CDWs in TbTe$_3$ during biaxial mechanical deformation between 250K and 375K. By directly measuring both CDW satellite intensities and resistivities along $\vec{a}$ and $\vec{c}$, we show that the CDW orientation continuously flips from $\vec{c}$ to $\vec{a}$ when the $a/c$ ratio is increased, with a coexistence of both orientations in the tetragonal region $a=c$. As those two parameters are linked to the gap, we can infer that the gap position changes from one band to another, and that both bands are gapped when $a=c$. Moreover, the transition temperature displays a linear behaviour with respect to $a/c$, with a minimum at $a=c$, which is well accounted for by a 2D tight-binding model of the system in the $\left(\vec{a}, \vec{c}\right)$ Te planes, with good quantitative match between theory and experiment. For the pure CDW along $a$ and CDW along $c$ phases, the gap saturates while $T_c$ does not, which questions the relationship between those quantities in RTe$_3$ systems. 
Our study opens great perspectives for the exploration of many electronic phase transitions in condensed matter systems with application of biaxial tensile stress at cryogenic temperatures.


\bibliography{biblio}

\end{document}


\appendix

\section{Resistivity measurements}
\subsection{Protocol}
\begin{itemize}
    \item set T to 350K, $F_a=0$ and $F_c=0$
    \item set working forces $F_a$ and $F_c$
    \item apply current (value) with Keithley 2611, measure voltage with nanovoltmeter 2182 in configuration $\rho_{aa}$ or $\rho_{cc}$
		\item procedure to switch from configuration $\rho_{aa}$ to $\rho_{cc}$
    \item cool down to 250K at rate 1K/min and measure U and I at (3 samples/K)
    \item heat up to 350K 
\end{itemize}

\subsection{Data treatment}
\begin{itemize}
    \item get $R_{aa}$ by doing U/I at each measurement point along $a$
    \item get $R_{cc}$ by doing U/I at each measurement point along $c$
    \item calculate $r=\frac{R_{aa}}{R_{cc}}$ and $x = A + B\ln{r} + C\left(\ln{r}\right)^2 + D\left(\ln{r}\right)^3 \dots$ with $A = 0.997229, B = 0.160697, C = 1.231448\cdot10^{-2}, D = -3.79740\cdot10^{-4}$.
    \item $H = \frac{\pi}{8}\sinh\left(\frac{\pi}{x}\right)$
    \item $\rho_{aa} = HeR_{aa}\frac{x}{l_a/l_c}$
    \item $\rho_{cc} = HeR_{aa}\frac{l_a/l_c}{x}$ with $l_a/l_c \sim 2.118$
    \item $A = \frac{\rho_{aa}}{\rho_{cc}}$
\end{itemize}

\subsection{Determination of $T_c$}

\section{Determination of lattice parameters and strain tensor}
\subsection{Orientation matrix and lattice parameters}
\subsection{CDW wavevector determination}
\subsection{Strain tensor}
$\tilde{\varepsilon}$: 

\begin{equation}
\tilde{\varepsilon} = 
\begin{pmatrix}
\varepsilon_{aa} & \varepsilon_{ab} & \varepsilon_{ac}\\
\varepsilon_{ba} & \varepsilon_{bb} & \varepsilon_{bc}\\
\varepsilon_{ca} & \varepsilon_{cb} & \varepsilon_{cc}
\end{pmatrix}
 = 
\begin{pmatrix}
\varepsilon_{aa} & \varepsilon_{ab} & \varepsilon_{ac}\\
\varepsilon_{ab} & 0 & \varepsilon_{bc}\\
\varepsilon_{ac} & \varepsilon_{bc} & \varepsilon_{cc}
\end{pmatrix}
\end{equation}

\begin{figure*}
\includegraphics[width = \textwidth]{Bragg_peaks_Diffabs}
\caption{\label{fig:bragg-pics}}
\end{figure*}

\begin{figure}
\includegraphics[width = \columnwidth]{lattice_parameters_Diffabs}
\caption{\label{fig:lattice-param}}
\end{figure}

\begin{figure*}
\includegraphics{resistivities_anisotropies}
\caption{\label{fig:resistivities_anisotropies}Use the figure* environment to get a wide
figure that spans the page in \texttt{twocolumn} formatting.}
\end{figure*}

\begin{figure}
\includegraphics[width = \columnwidth]{DTc_DRho}
\caption{\label{fig:DTc_DRho}}
\end{figure}

\begin{figure}
\includegraphics[width = \columnwidth]{satellite_peak_diffabs}
\caption{\label{fig:satellite-pic}}
\end{figure}

